\documentclass[conference]{IEEEtran}
\usepackage[pdftex]{graphicx}
\usepackage{enumitem}
\IEEEoverridecommandlockouts
\usepackage{cite}
\usepackage{amsmath,amssymb,amsfonts}
\usepackage{algorithmic}
\usepackage{graphicx}
\usepackage{textcomp}
\usepackage{xcolor}
\usepackage{cite}
\usepackage{algorithmic}
\usepackage[hyphens]{url}
\usepackage{hyperref}
\hypersetup{hidelinks}
\usepackage{cleveref}
\usepackage{algorithm}
\usepackage{gensymb}
\usepackage{booktabs}
\usepackage{float}
\usepackage{graphicx}
\usepackage{bm}
\usepackage{stmaryrd}
\usepackage{cancel}
\usepackage{mathtools}
\usepackage{subcaption}
\usepackage{caption}
\usepackage{todonotes}
\renewcommand{\baselinestretch}{0.99}

\def\BibTeX{{\rm B\kern-.05em{\sc i\kern-.025em b}\kern-.08em
    T\kern-.1667em\lower.7ex\hbox{E}\kern-.125emX}}

\begin{document}

\title{Enhancing power grid resilience to cyber-physical attacks using distributed retail electricity markets
\thanks{This work was supported by the US Department of Energy under Award DOE-OE0000920 and the Martin Society Fellowship for Sustainability from the MIT Environmental Solutions Initiative.}
}

% \author{\IEEEauthorblockN{Vineet Jagadeesan Nair}
% \IEEEauthorblockA{\textit{Dept. of Mechanical Engineering} \\
% \textit{Massachusetts Institute of Technology}\\
% Cambridge, MA, USA \\
% jvineet9@mit.edu}
% \and
% \IEEEauthorblockN{Priyank Srivastava}
% \IEEEauthorblockA{\textit{Dept. of Mechanical Engineering} \\
% \textit{Massachusetts Institute of Technology}\\
% Cambridge, MA, USA \\
% psrivast@mit.edu}
% \and
% \IEEEauthorblockN{Anuradha Annaswamy}
% \IEEEauthorblockA{\textit{Dept. of Mechanical Engineering} \\
% \textit{Massachusetts Institute of Technology}\\
% Cambridge, MA, USA \\
% aanna@mit.edu}
% }

\author{Vineet Jagadeesan Nair$^{1}$, Priyank Srivastava$^{2}$, and Anuradha Annaswamy$^{1}$
\thanks{$^{1}$Vineet Jagadeesan Nair and Anuradha Annaswamy are with the Department of Mechanical Engineering, Massachusetts Institute of Technology, Cambridge, MA 02139, USA. {\tt\small \{jvineet9, aanna\}@mit.edu}
$^{2}$Priyank Srivastava is with the Department of Electrical Engineering, Indian Institute of Technology Delhi, New Delhi -- 110016, India. {\tt\small psrivast@iitd.ac.in}
}
}

\maketitle

% To display page numbers - uncomment later
\thispagestyle{plain}
\pagestyle{plain}

\begin{abstract}
We propose using a hierarchical retail market structure to alert and dispatch resources to mitigate cyber-physical attacks on a distribution grid. We simulate attacks where a number of generation nodes in a distribution grid are attacked. We show that the market is able to successfully meet the shortfall between demand and supply by utilizing the flexibility of remaining resources while minimizing any extra power that needs to be imported from the main transmission grid. This includes utilizing upward flexibility or reserves of remaining online generators and some curtailment or shifting of flexible loads, which results in higher costs. Using price signals and market-based coordination, the grid operator can achieve its objectives without direct control over distributed energy resources and is able to accurately compensate prosumers for the grid support they provide.
\end{abstract}

\begin{IEEEkeywords}
Resilience, smart grid, electricity markets, optimization, cyber attacks
\end{IEEEkeywords}

\section{Introduction and motivation}
Cyber-physical attacks have become an increasingly prevalent threat to power systems operations around the globe. Recent events like the Ukraine power grid hacks in 2015 and 2016 \cite{case2016analysis}, and the Colonial pipeline ransomware attack in 2021 highlight the vulnerability of critical infrastructure to such events. Grid operators need to be able to quickly detect, isolate, and mitigate such attacks in order to maintain safe and reliable operations. At the same time, to curb emissions and reach net zero, the grid is also transitioning from fossil fuel-based thermal generation like coal and natural gas to renewable sources like wind and solar. Thus, unlike the current grid which is dominated by large centralized plants, the future grid will be highly decentralized. The penetration of distributed energy resources (DER) like rooftop solar photovoltaic (PV) systems, batteries, and electric vehicles is rapidly increasing, as well as flexible loads with demand response (DR) capabilities. A distributed grid also implies more endpoints of control at the edge, leading to more potential attack surfaces, which can make it more susceptible to cyber-physical attacks. We argue that by deploying a retail market that allows the insertion of operators at critical nodes in the distribution grid we can effectively monitor various nodes and coordinate them using market mechanisms. By doing so, we contend that we can actually enhance, rather than degrade resilience. 

Another major challenge is that, unlike conventional generators, grid operators generally do not have direct control over DERs which may have varied ownership and thus behave as autonomous agents. As these ownerships may be private or public, grid operators may not even have complete visibility over these resources in the distribution grid. This motivates the use of a market-based transactive framework \cite{Hao2017TransactiveResponse, Liu2017TransactiveImplementation}, which can incentivize various rational and trustable agents to provide grid services by varying their power injections in a desired manner, thus allowing efficient integration of DERs into the grid. These incentives can be in the form of market compensation for both active (P) and reactive power (Q) injection. The corresponding electricity price, which may vary both spatially and temporally, can then suitably compensate these agents for their grid services. Such a market structure would provide a high level of situational awareness (SA) to the distribution network operator, improve its ability to respond to attacks and strengthen grid resilience. It would also allow them to isolate attacks and mitigate these locally as much as possible and minimize reliance on the larger transmission grid, thereby preventing attacks in one region from affecting other areas. This can help avoid more widespread disruption and cascading failures. 

The rest of the paper is structured as follows: We provide a brief overview of prior works related to grid cybersecurity in \cref{sec:background}, and state our key contributions. The overall market structure and attack mitigation scheme are described in \cref{sec:method}, followed by a discussion on simulation results in \cref{sec:results}. Finally, we summarize our findings and outline several areas for future work and possible extensions in \cref{sec:conc}. Some further details are provided in \cref{sec:appendix}.

\section{Background and contributions \label{sec:background}}

\subsection{Prior work, literature review}
%\begin{itemize}
%\item connect with ukraine attack description and talk about some studies pointing what went wrong
%\item works (in control community or otherwise) addressing different aspects of the attack
%\item gap in existing literature, majority focus on security
%\item Some works on resilience 
%\item connection with our contributions -- no work utilizing local generation for resilience
%\end{itemize}

Cyberattacks could broadly be classified into disclosure, deception, or disruption \cite{DIBAJI2019394,chong19}. These attacks are respectively, the result of a confidentiality breach (allowing unauthorized users to access sensitive information), an integrity breach (intentional modification of signals from their true value), or an availability breach (blocking authorized users from accessing information). 
%Deception and disruption attacks cause direct harm to the system, while disclosure attacks are often executed to help launch deception and disclosure attacks later. 
Oftentimes, attackers use a combination of these attacks to inflict more damage to the system.
For example, the Ukraine power grid attack cite which rendered about 225,000 households without power \cite{case2016analysis} consisted of all three attack vectors. Yet another large-scale attack that has been explored in the context of power grids is MadIoT (Manipulation of Demand in IoT networks) \cite{soltan2018blackiot_new,shekari2022madiot_new}, where it is shown that a coordinated attack on high-wattage IoT devices could result in large-scale power blackouts.

% dynamic watermarking
% Not all is dark/gloomy paper
% Other 

Owing to the importance of protecting critical infrastructure against any cyberattacks, there has been a significant amount of work in the design of attack as well as defense methodologies \cite{Nguyen2020ElectricArt}. 
%The literature on defense against attacks could be broadly divided into prevention, resilience, and detection and isolation. 
The majority of the cybersecurity tools such as randomization \cite{mo2017privacy,nozari2017differentially}, cryptography \cite{farokhi2017secure,PramodT.C.2015}, and multiple-factor authentication \cite{Ometov18,Ibrokhimov19}, aim to prevent cyberattacks.
%In the context of power systems, the work cite(Xie et al 11) showed that under certain assumptions on the network topology, false data injection attacks could result in financial loss to the operators and consumers.
Since it is not always possible to protect the system against attacks, we focus on making the system resilient, i.e.,
minimizing the impact of malicious attacks and ensuring that the system keeps providing critical services even under stress \cite{bie17}.
Notable resilience approaches to deal with deception attacks include mean subsequence-reduced (MSR) algorithms and trust-based approaches.
MSR-based approaches filter out the suspicious values in updating the controller \cite{leblanc2017resilient,dibaji2017resilient,sundaram2016secure}; trust-based approaches are based on having redundant resources and making sure a subset of resources can always be trusted~\cite{momani2010survey,abbas2017improving}.
For example, in the context of power systems, the work~\cite{kim2011} proposes a trust-based approach, which protects a subset of measurements being used in state estimation in power systems against false data injection attacks. 

Despite these varied methods proposed in the literature, there are few approaches well-suited to large-scale attacks such as the Ukraine attack and MadIoT %\cite{soltan2018blackiot_new,shekari2022madiot_new} 
attacks. Even works that propose comprehensive approaches for preventive, protective, and emergency response rely on centralized visibility and control over generators, grid devices, and loads \cite{Huang2017IntegrationEnhancement,Soltan2020ProtectingAttacks}. The limited works connecting markets to resilience are largely focused on transmission, do not describe implementation details, or rely solely on existing market structures that are inadequate for the future DER-rich grid~\cite{Chen2020TowardOperators}.

% As evident from the recent attacks on the grid (cite ukraine, some others) and simulation studies on attacks using Internet of Things (IoT) botnets, like MaDIoT \cite{shekari2022madiot_new} and BlackIoT , it is imperative that alongside works on cybersecurity, the resilience of the system should also be increased.

% Resilience & markets
% Check if any market-based / transactive approaches to resilence/cyber-physical attacks

% In particular, we employ a trust-based approach, where the trust values of the available resources are updated using the network traffic and ...(cite partha). 
% one sentence on recovery (detection and isolation) needed...

\subsection{Our contributions}

In scenarios where several nodes are simultaneously compromised, it is not always possible to have a centralized decision-maker solely responsible for mitigating the attack. A more extensive decision-making that is distributed in nature is called for. Operators at multiple levels of hierarchy that (i)~have visibility over different sets of agents, (ii) are judiciously located over the large-scale system, (iii) utilize this visibility to identify the attack, and (iv) use their situational awareness to mitigate the impacts are essential. Our contributions are summarized below:
% The contribution of this paper is the proposal of a market framework to determine these operators and how situational awareness (SA) can be made available to the operators. We propose that this market has a hierarchical structure, and collocate the operators with the electrical structure of the distribution grid which includes a primary feeder and a secondary feeder. The corresponding primary and secondary markets provide their operators with SA and commitment scores that describe the reliability of the agents in following the market settlements during operation. 
\begin{itemize}
\item We propose a market-based framework with a hierarchical structure to provide oversight over possible attacks and to coordinate resources in a distributed manner. To our knowledge, ours is the first paper that leverages a market-based, transactive energy approach to detect, isolate, and mitigate cyber-physical attacks.
\item A two-tier structure with a primary market and secondary market provides the operators with visibility over their corresponding agents.
\item A novel commitment score metric is generated for each market agent which allows operators to determine how reliable assets are in responding to market signals
(schedules and prices) which in turn is used to identify agents that can ensure grid resilience.
\item We develop a method for market operators to re-dispatch resources via fully distributed, privacy-preserving optimization, by tuning the objective function parameters. 
\item This framework also provides flexibility to balance different priorities, like maximizing local social welfare versus global grid objectives such as minimizing transmission imports, while ensuring all constraints related to power flow, budget balance, and capacity limits are satisfied.
% \item Suitable cost functions and constraints are proposed for the operation of the primary and secondary market, ensuring that grid constraints of power balance with nonlinearities and constraints due to capacity limits are all satisfied. A distributed optimization framework is utilized that ensures privacy of the market participants. Such a framework allows flexibility in balancing different priorities,
% Such as maximizing local social welfare versus global grid objectives such as …. %fill this-in
% \item In contrast to other papers, we employ a market structure in order to 
% \item Our hierarchical approach allows for attack recovery (detection, isolation, and mitigation) even when the grid operators don't have full visibility over the distribution grid and market participants. 
% \item A novel commitment score metric allows us to track how reliable assets are in responding to market signals (schedules and prices) and enables a trust-based approach to ensure attack recovery.
\end{itemize}

\subsection{Relation to our previous work \label{sec:relations}}

We build upon our prior work in \cite{Nair2022AEdge} and leverage the market structure developed previously for this entirely new application of mitigating cyber-physical attacks to increase resilience. The optimization problem formulations have been adapted from the previous work and are thus similar, but have a few modifications and improvements. For instance, the commitment score term in \cref{eq:sm_cost1} quantifies the aggregate reliability based on the \textit{flexibility} in net injections provided by the SMAs (in terms of deviations from their baseline), rather than just the actual values of the injections themselves. To ensure feasibility of the SM optimization, we also enforce a relaxed, integral version of the budget balance constraints in \cref{eq:budgetP,eq:budgetQ} that requires the SMO to break even only over a longer time horizon, rather than over every single PM time step. We also included and interpreted results from the SM simulation showing the evolution of our commitment score metric over time. For the distributed optimization in the PM in \cref{sec:dist_opt}, we deploy the modified NST-PAC algorithm instead which is faster and more private, compared to the PAC algorithm which was used in the previous work. We also develop a detection and mitigation approach for disruption attacks, provide an intuitive justification for our algorithm, and validate it using simulations of two distinct attack scenarios. 

\section{Methodology \label{sec:method}}

\begin{table}[]
\centering
\begin{tabular}{@{}ll@{}}
\toprule
\textbf{Acronym} & \textbf{Definition}               \\ \midrule
PM/SM            & Primary/Secondary market          \\
PMO/SMO          & Primary/Secondary market operator \\
SMA              & Secondary market agent            \\
WEM              & Wholesale energy market           \\
DSO              & Distribution system operator      \\
PCC              & Point of common coupling \\ \bottomrule
\end{tabular}
\caption{Key market-related acronyms. \label{tab:acronyms}}
\end{table}

\subsection{Problem statement and market-based solution}

Deception and disruption attacks on a distribution grid %are events that 
can significantly alter the nodal power injections throughout the network. 
We focus specifically on attacks that decrease the net generation available at multiple nodes simultaneously. This could be due to either physical attacks on generators or grid infrastructure or cyber-attacks that affect control systems and/or actuators. Such cyber-attacks generally correspond to disruption attacks that breach availability \cite{DIBAJI2019394}. Disruptions can occur in several different ways. Attackers can hack individual controllers to alter generation output, e.g. causing solar smart inverters to curtail more power. On a larger scale, control centers can be attacked to disconnect nodes from the network and create outages, as in Denial of Service (DoS) attacks. Attackers can also hack high-wattage devices to manipulate and increase electricity consumption. In addition to disruption attacks, certain types of deception attacks that breach data integrity can also result in similar changes in generation or load, e.g. by tampering with sensor measurements used by controllers or by directly modifying control signals and command setpoints. Such generation and/or load-altering attacks would eventually lead to a drop in net total generation available in the network. The distribution grid operator then faces the challenge of meeting this shortfall between demand and supply, while minimizing the amount of excess power imported from the transmission system. We propose to solve this problem using a hierarchical retail electricity market structure \cite{Nair2022AEdge} that provides situational awareness, measures the commitment reliability of market agents, and optimally coordinates and dispatches the distributed energy resources for attack mitigation. 

\begin{figure}
\centering
     \includegraphics[width=0.9\columnwidth]{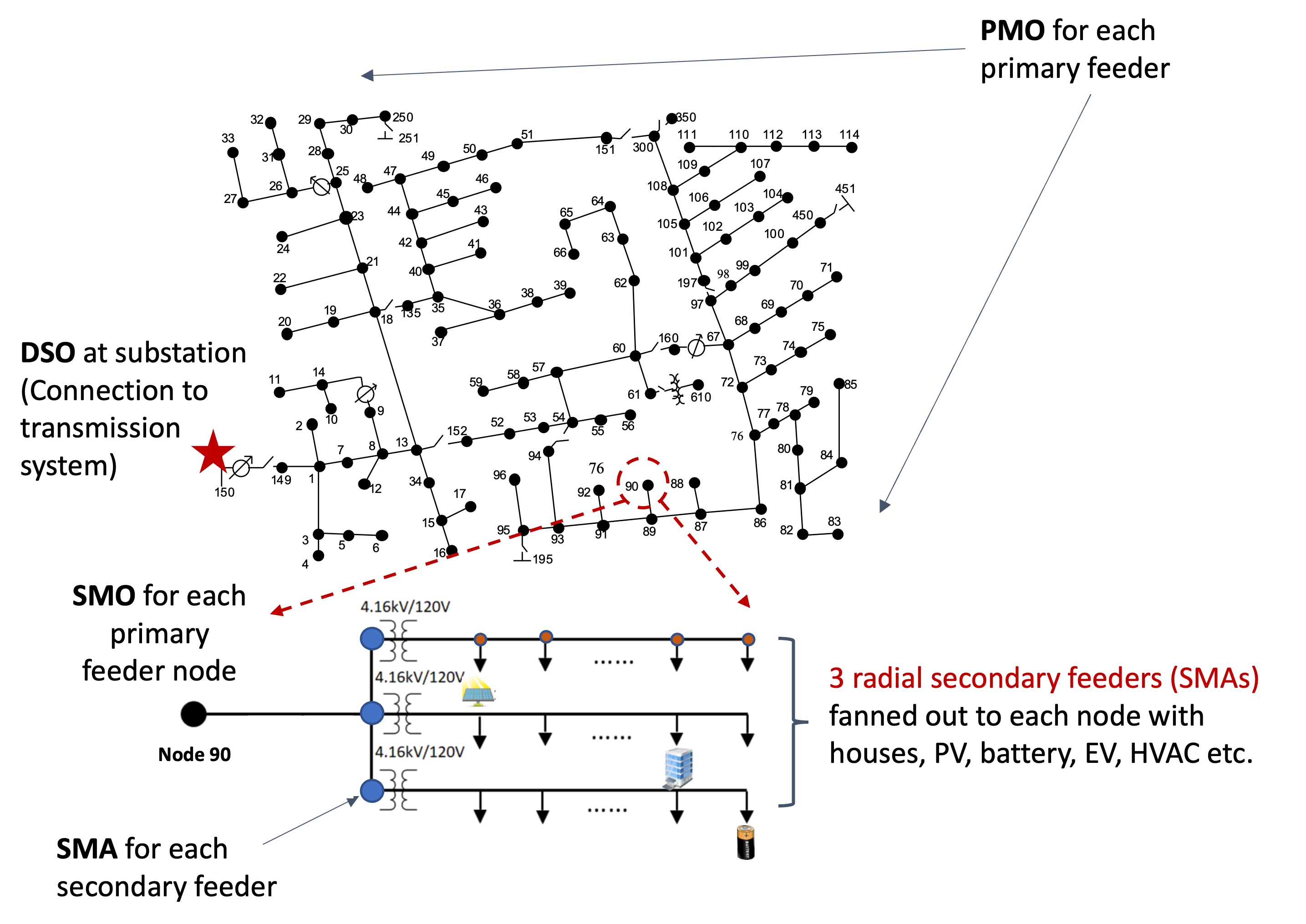}
     \caption{Hierarchical market co-located with the distribution grid, showing a primary feeder network based on the IEEE-123 node test case. This figure has been adapted from \cite{Nair2022AEdge}. \label{fig:network_diagram}}
\end{figure}
\begin{figure}
\centering
\includegraphics[width=0.9\columnwidth]{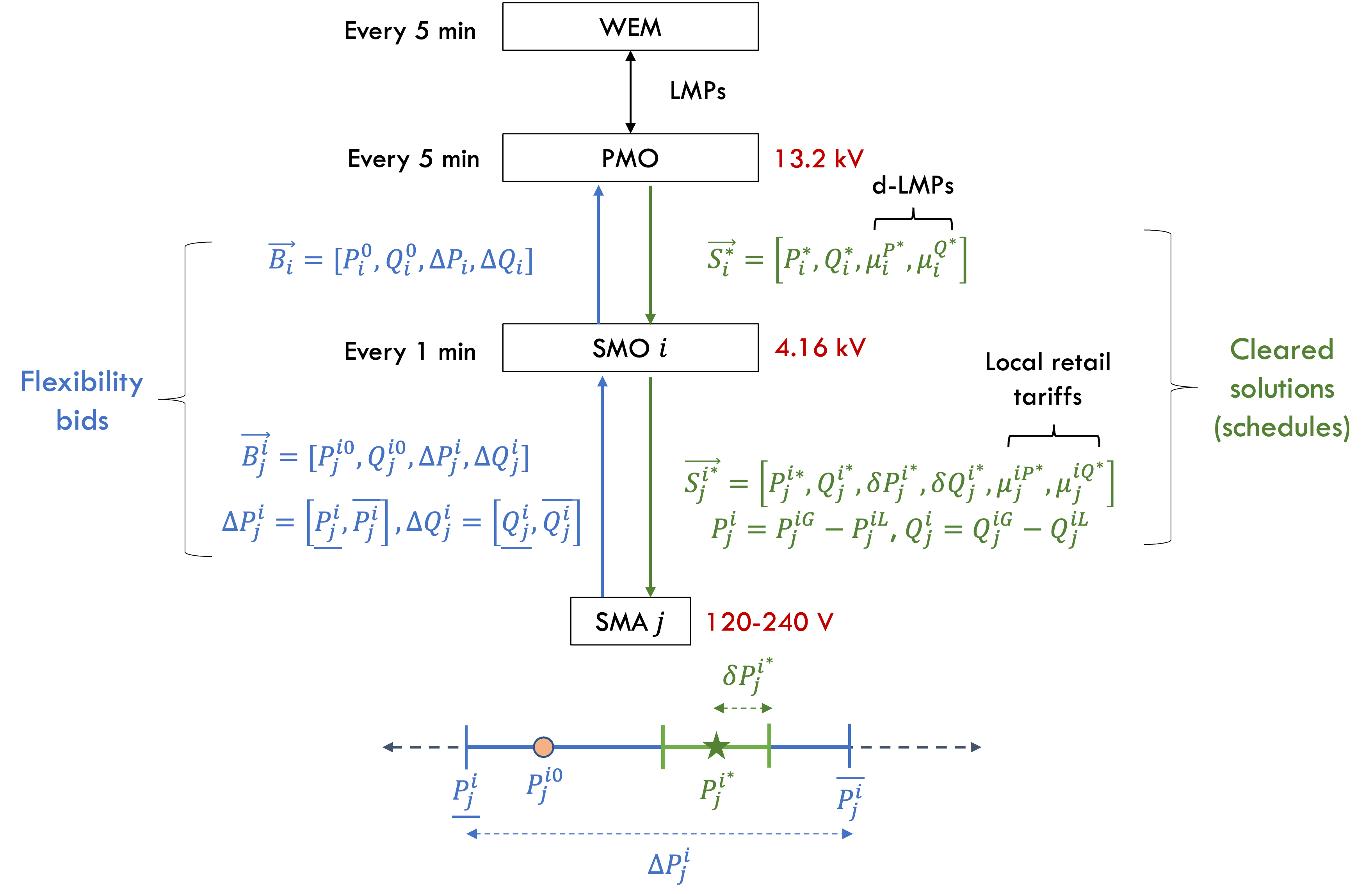}
\caption{Market inputs and outputs, adapted from \cite{Nair2022AEdge}. \label{fig:in-out}}
\end{figure}

The market is collocated with the electrical structure of the distribution grid, as shown in \cref{fig:network_diagram}. This consists of an upper-level primary market (PM) at the medium-voltage level, where the primary market operator (PMO) oversees multiple secondary market operators (SMO), which in turn coordinates several agents in the secondary market (SM) downstream at the low-voltage level. Each SMO represents a single primary feeder node in the distribution network while a secondary market agent (SMA) represents a secondary feeder. The PMO at the substation is connected to the main transmission grid and thus also participates in the wholesale energy market (WEM). Both the SM and the PM are overseen by a proposed distribution system operator (DSO) which may interact with existing entities like an independent system operator (ISO) or a regional transmission operator (RTO).
% \begin{figure}
% \centering
% \includegraphics[width=0.35\columnwidth]{flowchart_v2.png}
% \caption{Market participants. \label{fig:agents}}
% \end{figure}
The inputs and outputs for the primary and secondary markets are summarized in \cref{fig:in-out}. The starting points for each market are bids submitted by each agent, consisting of a baseline power injection for active and reactive power ($P^0, Q^0$) along with associated upward and downward flexibility limits from that baseline. Both the SM and PM clearing result in corresponding prices and power setpoints (or schedules) which are used to establish bilateral contracts between the market operators and the market participants (or agents). Note that several equations presented in the following sections \ref{sec:sm} to \ref{sec:pm} have been replicated from \cite{Nair2022AEdge} with some modifications as noted in \cref{sec:relations}.

%Note that for this work, we focus only on the primary and secondary markets at the medium and low-voltage levels respectively. We will develop the consumer-level market closest to the end-users, as part of future work.
\subsection{Secondary market (SM) model \label{sec:sm}}
The SM is cleared using decentralized multiobjective optimization, wherein each SMO is responsible for scheduling multiple SMAs $j \in \mathcal{N}$ connected to it below. In a sense, the SMO can also be thought of as an aggregator that coordinates its SMAs to optimally participate in the primary, and eventually the wholesale markets, upstream. The SMO optimizes for different objectives (listed in decreasing order of relative importance): (i) maximizing aggregate reliability ($f_j^1$) by utilizing more flexibility from SMAs with higher commitment scores (defined in the next section), (ii) minimizing net costs ($f_j^2$), (iii) maximizing the total flexibility from its SMA schedules ($f_j^3$), and (iv) minimizing the disutility of SMAs associated with providing flexibility ($f_j^4$). The constraints include operational capacity limits (\cref{eq:Plim,eq:Qlim}), a ceiling on retail tariffs for compensating or charging SMAs (\cref{eq:fairness}), power balance between the upper-level PM and lower-level SM (\cref{eq:PQbalance}), and budget balance to ensure that the SMO breaks even over a specified time horizon (\cref{eq:budgetP,eq:budgetQ}). The PM and SM timesteps are denoted by $t_p$ and $t_s$ respectively. We assume here that the PM is cleared along with the WEM every $\Delta t_p = 5 \; min$ and the SM is cleared more frequently every $\Delta t_s = 1 \; min$.

For each SMA $j$, the SM optimization solves for the active and reactive power setpoints ($P^*_j, Q_j^*$) as well as a flexible range around them defined by $\delta P_j^*, \delta Q_j^*$, i.e. the SMO directs the SMAs to stay within $[P_j^* -\delta P_j^*, P_j^* + \delta P_j^*]$. The overall vector of decision variables is $\vec{S}_j = [P_j, Q_j, \delta P_j, \delta Q_j, \mu^P_j, \mu^Q_j],  $ The clearing also sets the retail tariffs $\mu_j^{P^*}, \; \mu_j^{Q^*}$. The optimization problem is given below in \cref{eq:opt}, where $\beta^P_j, \; \beta^Q_j$ are the disutility coefficients associated with flexibility and $\mathcal{N}$ is the set of all SMAs under the given SMO. Note that throughout this work, $P = P_G - P_L, \; Q = Q_G - Q_L$ denote \textit{net} injections, i.e., generation minus load.
\begin{subequations}
\label{eq:opt}
\begin{align}
& \min_{\vec{S}_j} \sum_{j \in \mathcal{N}} \{f_j^1,f_j^2,f_j^3,f_j^4\} \label{eq:cost} \\
& f_j^1 \succ f_j^2 \succ f_j^3 \succ f_j^4 \label{eq:ranking} \\ 
& f_j^1 = -C_j ((P_j - P_j^{0})^2 +  (Q_j - Q_j^{0})^2) \;  \label{eq:sm_cost1} \\
& f_j^2  = \mu_j^{P} P_j + \mu_j^{Q} Q_j, \; f_j^3 = -(\delta P_j + \delta Q_j) \label{eq:sm_cost2} \\
& f_j^4 = \beta_j^{P}(P_j - P_j^{0})^2 + \beta_j^{Q}(Q_j - Q_j^{0})^2 \label{eq:sm_cost3} \\
& \text{subject to:} \nonumber\\
& \underline{P}_j + \delta P_j \leq P_j \leq \overline{P}_j - \delta P_j\label{eq:Plim} \\ 
& \underline{Q}_j + \delta Q_j\leq Q_j \leq \overline{Q}_j - \delta Q_j\label{eq:Qlim} \\ 
& \delta P_j, \delta Q_j \geq 0, \; 0 \leq \mu_j^{P} \leq \overline{\mu}^{P}, \; 0 \leq \mu_j^{P} \leq \overline{\mu}^{Q}  \; \label{eq:fairness} \\  % \label{eq:flex}
& \sum_{t_p}\sum_{t_s} \sum_{j \in \mathcal{N}} \mu_j^{P}(t) P_j(t) \Delta t_s \leq \sum_{t_p} \mu^{P^*}(\hat{t}_p) P^*(\hat{t}_p) \Delta t_p \label{eq:budgetP} \\
& \sum_{t_p}\sum_{t_s} \sum_{j \in \mathcal{N}} \mu_j^{Q}(t) Q_j(t) \Delta t_s \leq \sum_{t_p} \mu^{Q^*}(\hat{t}_p) Q^*(\hat{t}_p) \Delta t_p \label{eq:budgetQ}\\ 
& \sum_{j \in \mathcal{N}} P_j(t_s) = P^*(\hat{t}_p), \quad \sum_{j \in \mathcal{N}} Q_j(t_s) = Q^*(\hat{t}_p) \label{eq:PQbalance}
\end{align}
\end{subequations}
Such multiobjective problems generally do not elicit unique solutions, but we can solve for a set (or frontier) of Pareto-optimal points, and the SMO can choose amongst these according to their ranking or prioritization of objectives as in \cref{eq:ranking}. In general, these four objective terms may have very different magnitudes and it is quite challenging to normalize these appropriately \cite{Marler2004SurveyEngineering}. Thus, we instead use a hierarchical approach, details of which are in the appendix in \cref{sec:hierarchy_app}.

\subsubsection{Commitment scores \label{sec:commit}}
We formulate a commitment score that measures how reliable each agent is in following through with their contractual commitments. The SMO monitors the actual injections of different SMAs to update the score $C_j \in [0,1]$ for each of the SMAs $j$. Higher scores indicate greater commitment reliability. Further details of score computation are described in the appendix in \cref{sec:commit_app}. We note that to compute and update the commitment scores, the SMO needs to be able to track the actual responses of SMAs. This can be done through smart meters or other advanced metering infrastructure (AMI) deployed by the SMAs. It is assumed that the SMAs consent to this by participating in the SM. One can also argue that privacy risks in the SM are relatively lower, due to the smaller scale of the market relative to the PM. Depending on the time resolution of SMA monitoring data available to the SMO, we can also make score updates less frequently rather than at every secondary market clearing time $t_s$. Finally, we note that disruption and deception attacks could affect the ability of corrupted SMAs to follow through on their contractual commitments. This would result in a drop in their commitment scores and in turn, influence the optimization problem and future market clearings. This also allows the SMO to take into account the susceptibility of different SMAs to attacks while running the SM.

\begin{figure}
\centering
\includegraphics[width=0.7\columnwidth]{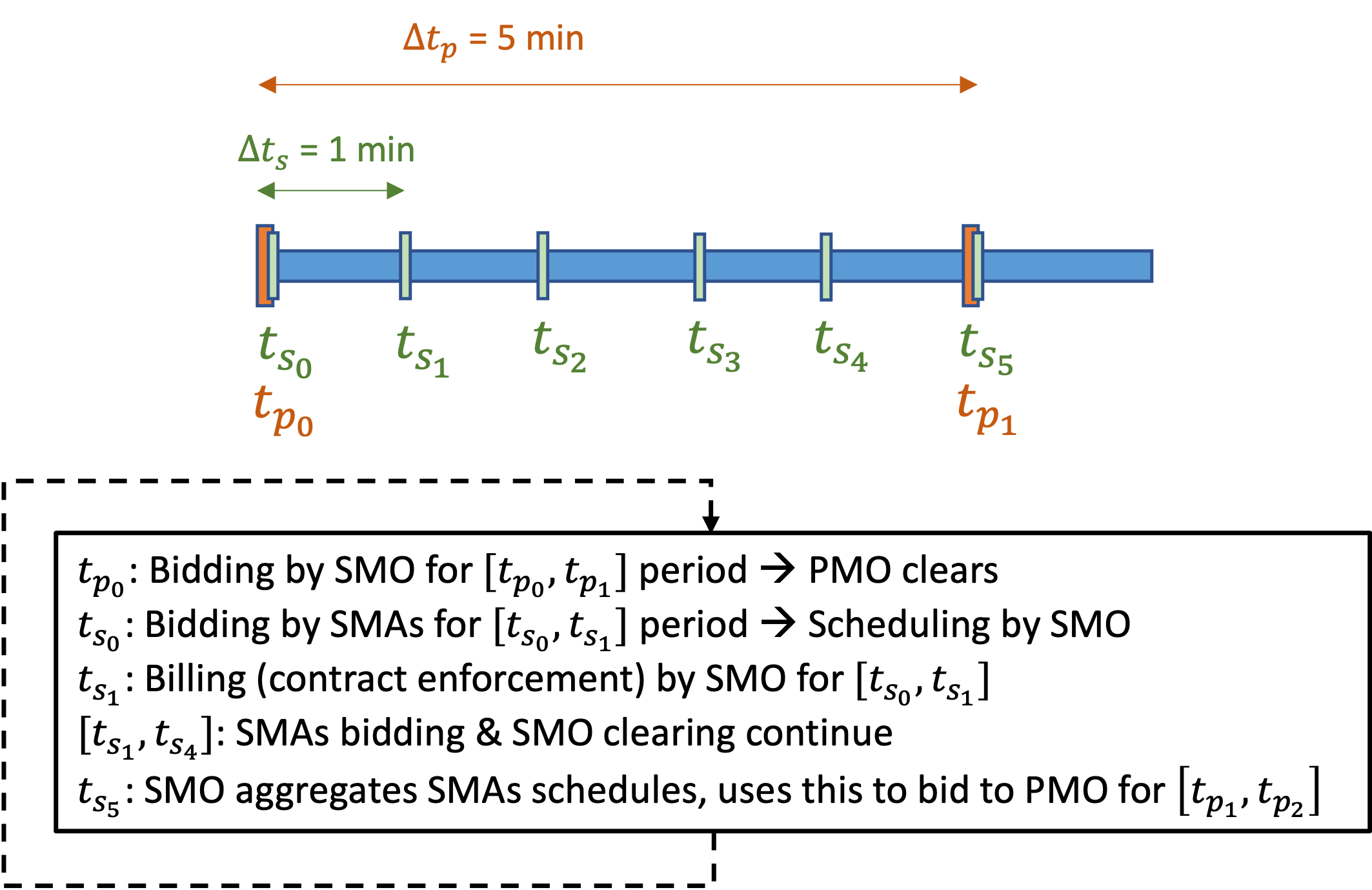}
\caption{Timeline of SM and PM operation. \label{fig:timeline}}
\end{figure}

\subsection{Primary market (PM) model \label{sec:pm}}
The PM clearing involves solving an alternating current optimal power flow (ACOPF) problem, assuming a radial, balanced distribution network topology. Here, we use the well-known branch flow or nonlinear DistFlow model \cite{Molzahn2017ASystems} which employs a second-order conic program (SOCP) relaxation for the quadratic equality constraint that defines the apparent power $S$. Other power flow constraints include Ohm's law, power balance, thermal limits on line power flows, voltage bounds, and active and reactive power injection limits for each SMO. This solves for the active and reactive power setpoints for each SMO $i$ located at primary feeder nodes. We can also derive the nodal prices for active and reactive power for each SMO by inspecting the dual variables associated with the corresponding power balance constraints. We refer to these are distribution-locational marginal prices (d-LMP). Further details of the PM model are described in the appendix in \cref{sec:pm_opf_app}.

\subsection{Distributed optimization \label{sec:dist_opt}}
For the SM clearing, a decentralized optimization approach is well-suited since each SM only involves the SMO and a small number of SMAs. However, the PM can potentially involve many more agents (SMOs), especially for large primary feeder networks. Thus, we employ a distributed optimization algorithm based on Proximal Atomic Coordination (PAC) in which each SMO communicates primarily with its neighboring SMOs \cite{Romvary2022AOptimization}. This allows us to decompose the global optimization problem and reduce the computational burden. In addition to making the problem more tractable and scalable for larger networks, this method helps preserve some privacy since SMOs don't have to share all their information with a centralized agent. This also helps reduce communication latencies relative to a purely centralized approach.

PAC is a distributed algorithm in which individual agents exchange a subset of primal and dual variables with neighboring nodes at each iteration. These coordination variables help satisfy coupled global network constraints (imposed by the grid power physics) while each SMO solves its local optimization problem. The iterative approach continues until the algorithm (provably) converges to globally feasible and optimal solutions for all the SMOs. We now briefly describe the details of this algorithm. For a given global optimization problem with equality and inequality constraints for $S$ number of nodes (or agents):
\begin{equation}
    \min _x \sum_{i=1}^S f_i(x) \text { s.t. } G x=b, \quad H x \leq d 
\end{equation} 
We can decompose this into $\mathcal{S} = \{1,2,\dots S\}$ coupled optimization problems, known as atoms (representing each SMO $i$). We separate the vector of all decision variables $x$ into two sets: $\mathcal{L} = \{L_i, \forall i \in \mathcal{S}\}$ and $\mathcal{O} = \{O_i, \forall i \in \mathcal{S}\}$ which is a partition of decision variables into those that are owned and copied by atom $i$, respectively. We can similarly also decompose the constraints into sets owned by each atom $\mathcal{C} = \{C_i, \forall i \in \mathcal{S}\}$. These variable copies can then be used to satisfy coupled constraints and global objectives. 

The decomposed (or atomized) optimization problem is shown in \cref{eq:opt_atom}, where $a_j$ and $f_j(a_j)$ are the primal decision variables (both owned and copies) and individual objective functions corresponding to each SMO atom, respectively. $G_j$ and $H_j$ are the atomic constraint submatrices of $G$ and $H$, while $b_j$ and $d_j$ are subvectors of $b$ and $d$ of the right hand side constraint vectors $b$ and $d$, respectively. $B$ is the directed graph incidence matrix defining the owned and copied atomic variables. This incidence matrix allows us to fully parallelize the distributed optimizing by defining coordination or consensus constraints, which enforce that all the copied variables for each atom $j$ must equal the values of their corresponding owned values in every other atom $i \neq j$. $B_j$ and $B^j$ denote the incoming and outgoing edges for atom $j$ respectively.
\begin{align}
& \min _{a_j} \sum_{j \in K} f_j\left(a_j\right) \label{eq:opt_atom} \\
& \text { s.t. } G_j a_j=b_j, \; H_j a_j \leq d_j, \; B_j a=0 \; \forall j \in K  \nonumber\\
& B_{im} \triangleq \begin{cases}
-1, & \text { if } i \text { is 'owned" and } m \text { a related "copy" } \\
1, & \text { if } m \text { is "owned" and } i \text { a related "copy" } \\
0, & \text { otherwise } \end{cases} \nonumber
\end{align}
% \begin{equation}
%  B_{im} \triangleq \begin{cases}
% -1, & \text { if } i \text { is 'owned" and } m \text { a related "copy" } \\
% 1, & \text { if } m \text { is "owned" and } i \text { a related "copy" } \\
% 0, & \text { otherwise } \end{cases}
% \end{equation}
This work employs an enhanced, accelerated version called NST-PAC developed in \cite{Ferro2022AHubs}. It is a primal-dual method incorporating both $L2$ and proximal regularization terms. The convergence speed is increased by using time-varying gains and Nesterov-accelerated gradient updates for both primals and duals. The augmented Lagrangian is first atomized or decomposed for each node or SMO, introducing dual variables $\eta$ and $\nu$ corresponding to the (primal) equality and coordination (or consensus) constraints respectively:
\begin{align}
\mathcal{L}(a, \eta, \nu) & =\sum_{j \in K}\left[f_j\left(a_j\right)+\eta_j^T\left(G_j a_j-b_j\right)+\nu_j^T B_j a\right] \nonumber\\
& =\sum_{j \in K}\left[f_j\left(a_j\right)+\eta_j^T\left(G_j a_j-b_j\right)+\nu^T B^j a_j\right] \nonumber\\
& \triangleq \sum_{j \in K} \mathcal{L}_j\left(a_j, \eta_j, \nu\right)
\end{align}
The iterative NST-PAC algorithm consists of the following steps at each iteration $\tau$:
\begin{align}
a_j[\tau+1] & = \underset{a_j}{\operatorname{argmin}}\big\{\mathcal{L}_j\left(a_j, \hat{\eta}_j[\tau], \hat{\nu}[\tau]\right)\\
& +\frac{\rho_j \gamma_j}{2}\left\|G_j a_j-b_j\right\|_2^2 + \frac{\rho_j \gamma_j}{2}\left\|B_j a_j\right\|_2^2 \nonumber\\
& +\frac{1}{2 \rho_j}\left\|a_j-a_j[\tau]\right\|_2^2\big\} \nonumber\\
 \hat{a}_j[\tau+1] &= a_j[\tau+1]+\alpha_j[\tau+1]\left(a_j[\tau+1]-a_j[\tau]\right) \nonumber\\
 \eta_j[\tau+1]& = \hat{\eta}_j[\tau]+\rho_j \gamma_j\left(G_j \hat{a}_j[\tau+1]-b_j\right) \nonumber\\
 \hat{\eta}_j[\tau+1]&= \eta_j[\tau+1]+\phi_j[\tau+1]\left(\eta_j[\tau+1]-\eta_j[\tau]\right) \nonumber\\
 \text {Communicate } & \hat{a}_j \text { for all } j \in[K] \text { with neighbors } \nonumber\\
 \nu_j[\tau+1]&= \hat{\nu}_j[\tau]+\rho_j \gamma_j B_j \hat{a}_j[\tau+1] \nonumber\\
 \hat{\nu}_j[\tau+1]&= \nu_j[\tau+1]+\theta_j[\tau+1]\left(\nu_j[\tau+1]-\nu_j[\tau]\right)\nonumber \\
 \text {Communicate } & \hat{\nu}_j \text { for all } j \in[K] \text { with neighbors } \nonumber
\end{align}
The algorithm further protects privacy by masking both the primal and dual variables. Masking is implemented by using iteration-varying and atom-specific parameters $\alpha_j[\tau]$, $\phi_j[\tau]$ and $\theta_j[\tau]$. Masking the dual variables (or shadow prices), in particular, is desirable since these may reveal sensitive data related to costs, operating constraints, or other preferences of SMOs. Instead, masked variables $\hat{a}$ and $\hat{\nu}$ are exchanged between atoms. By iteratively solving the local, decomposed optimization problems across all SMOs, NST-PAC (and PAC) provably converge to the globally optimal ACOPF (relaxed) solutions for the whole primary feeder \cite{Romvary2022AOptimization,Ferro2022AHubs}.

\subsection{Attack detection and mitigation}

Having described the detailed SM and PM formulation, we now leverage the market mechanism to effectively mitigate cyber-physical attacks on the distribution grid. Our approach uses the distributed market to coordinate resources using market or price-based transactive control while minimizing the amount of communication needed. For this study, we focus on disruption or DoS attacks. These can be due to either crashed faults where the nodes are completely taken offline, or when the devices are artificially manipulated or corrupted to alter their power injected into (in case of generation) or drawn from (in case of load) the network.
%, but we can easily generalize to deception attacks where power values at nodes are artificially manipulated.

The PMO is located at the substation or PCC connected with the transmission grid. It constantly monitors the net total power that the primary feeder imports (exports) from (to) the main grid. Thus, when resources within the feeder (generators or loads) are attacked, this will show up as a noticeable change in the net injection at the PCC, $P_{PCC}$. When this change is above some predefined threshold, the PMO will flag the event as an attack and initiate the mitigation efforts. Our mitigation approach does not require the PMO to have real-time visibility over the actual power injections at each SMO. While our hierarchical coordination helps to detect if there is an attack and determine its magnitude and impact, we do not focus on precisely identifying the attacked resource(s) or node(s) since this is not required for our mitigation approach. This helps improve robustness since the PMO does not rely on any information sent by the SMOs themselves - these communication channels may be corrupted or even non-functional during an attack. In a sense, we can argue that this fact also makes the approach less vulnerable to potentially malicious or compromised SMO agents who may send the PMO incorrect signals. The PMO does not have direct control over any of the SMOs or SMAs. However, it can achieve distributed coordination by cleverly modifying coefficients in the objective function and broadcasting these updates to all the SMAs. To illustrate how the PMO can derive the right coefficient updates, let us consider the objective function in \cref{eq:cost:1} for the PM ACOPF problem. For ease of exposition, this can be rewritten in a simplified manner as follows:
\begin{gather}
    \sum_{i=1}^n \left(\frac{1}{2} \; \alpha_i P_i^{G^2} + \beta_i \left(P_i^L - P^{L0}_i\right)^2\right) + \xi \cdot losses \label{eq:update_obj}\\
    \alpha_i^{\prime} = \Delta_{\alpha} \alpha_i, \; \beta_i^{\prime} = \Delta_{\beta} \beta_i, \; \xi^{\prime} = \Delta_{\xi} \xi; \; \alpha, \beta, \xi, \Delta > 0 \label{eq:update1}\\
    \Delta_{\alpha} = \Delta_{\beta} = \frac{|P_{PCC}|}{|P_{PCC}^{\prime}|}, \; \Delta_{\xi} = \frac{|P_{PCC}^{\prime}|}{|P_{PCC}|} \label{eq:update2}
    % \alpha, \beta, \xi, \Delta > 0 \nonumber
\end{gather}
To derive the coefficient updates, the PMO compares the net total load from the entire primary feeder at the substation (PCC) before and after the attack, which is given by $P_{PCC}$ and $P_{PCC}^{\prime}$, respectively. We can then compute coefficient updates based on the ratio between these two values. Updating the coefficients as in \cref{eq:update1,eq:update2} results in a re-dispatch of the SMOs in the PM that successfully mitigates the attack. In our context, successful attack mitigation means that the market operator can meet the shortfall between demand and supply by leveraging the flexibility of local resources within the feeder as much as possible, to minimize its reliance on the main transmission grid. By isolating and locally mitigating the attack in the distribution network, we can prevent it from impacting the rest of the grid, which may otherwise cause more serious issues for grid frequency and stability.

\begin{figure}
\centering
\includegraphics[width=0.85\columnwidth]{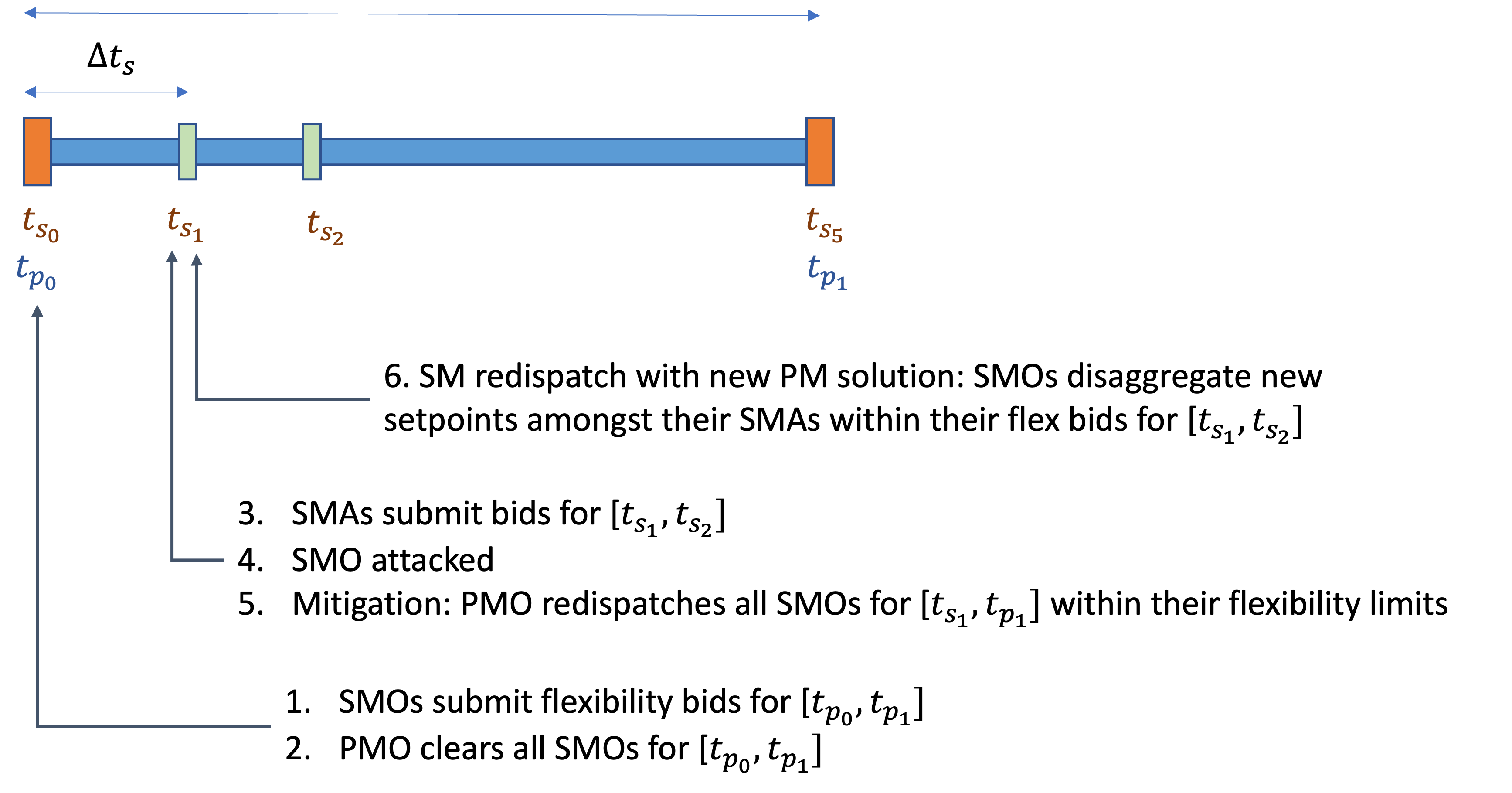}
\caption{Timeline of attack detection and mitigation. \label{fig:attack_timeline}}
\end{figure}

We now provide an intuitive explanation for why the above coefficient updates result in the desired behavior, by re-weighting the three terms in the objective function in \cref{eq:update_obj}. Without loss of generality, let us consider attack scenarios where several distributed local generator SMOs are attacked resulting in an increase in net feeder load, i.e. $|P_{PCC}^{\prime}| > |P_{PCC}|$ (note that both $P_{PCC}, P_{PCC}^{\prime} < 0$ since net loads are negative injections). This results in the following updates:
\begin{enumerate}
    \item $\Delta_{\alpha} < 1$ which artificially lowers cost coefficients $\alpha_i$ to dispatch more local generation from remaining online SMOs instead of importing power from the WEM. Here, we utilize the upward flexibility provided by SMOs bidding into the PM.
    \item $\Delta_{\beta} < 1$ which reduces the disutility coefficients $\beta_i$ to encourage more demand response via load shifting and/or curtailment, by utilizing the downward flexibility provided by SMOs bidding into the PM.
    \item $\Delta_{\xi} > 1$ which penalizes electrical line losses more heavily in the objective function, which discourages imports from the transmission grid in favor of dispatching more local DERs. This is because distribution grids have higher resistance to reactance ($\frac{R}{X}$) ratios and are more lossy. Thus, prioritizing the loss minimization objective makes it more efficient to utilize local generation closer to the loads being served.
\end{enumerate}
A more rigorous derivation of these update factors will be provided as part of future work. After deriving the multiplicative coefficient update factors $\Delta_{\alpha}, \Delta_{\beta}, \Delta_{\xi}$, the PMO can broadcast these common values to all the SMOs simultaneously. The PM is then re-dispatched via NST-PAC optimization using the new objective functions for each SMO. This results in new P and Q setpoints for SMOs, along with new nodal d-LMPs. This is followed by each SMO also re-dispatching their SM, to disaggregate the new setpoints among their SMAs. A timeline of the key events is shown in \cref{fig:attack_timeline}.
\subsection{Communication infrastructure}
\begin{figure}
\centering
\includegraphics[width=0.5\columnwidth]{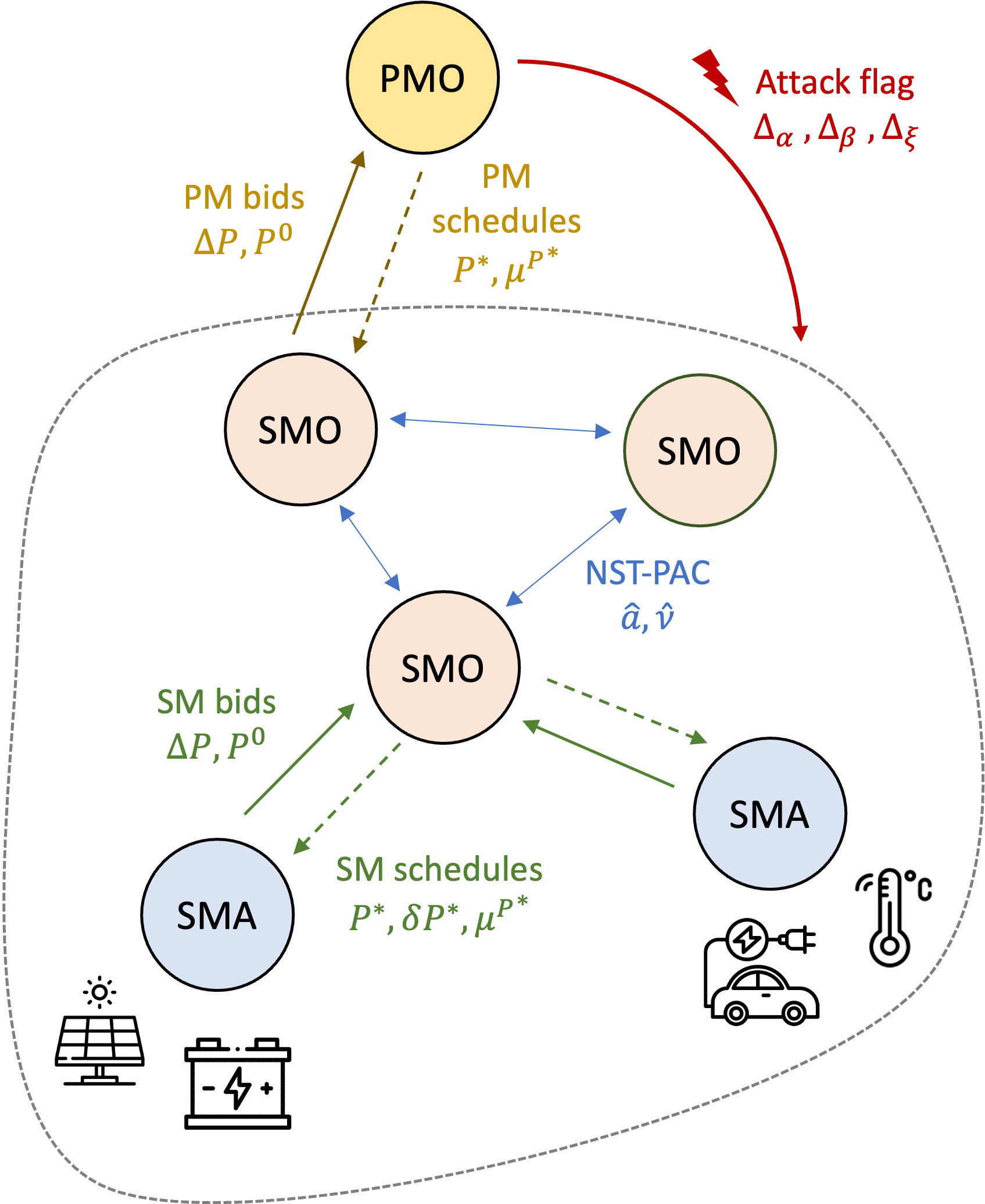}
\caption{Communication among market participants during both nominal and attack scenarios. \label{fig:comm_scheme}}
\end{figure}

The communication steps among different participants in the SM and PM are shown in \cref{fig:comm_scheme}. There is peer-to-peer (P2P) communication directly between SMOs to facilitate the exchange of masked primal-dual variables during the negotiation iterations of the NST-PAC algorithm. This can be implemented via either a synchronous or asynchronous request-reply communication scheme. A one-to-many publish-subscribe (pub/sub) is used by the PMO to communicate with all the SMOs, as well as by the SMO to communicate with its SMAs \cite{Lehnhoff2012OPCGrids}. This is used to exchange information about market bids and schedules, and thus establish bilateral contracts between the market operators and their agents. When an attack occurs, the PMO also uses this scheme to flag the attack and broadcast the objective function coefficient updates. Our attack mitigation approach only requires one-way communication from the PMO to all SMOs since communication from the SMO to the PMOs may be unavailable under an attack.
% Synchronous or asynchronous request-reply communication amongst the SMOs, peer-to-peer (P2P)

\section{Results and discussion \label{sec:results}}

\subsection{Numerical simulation setup}
We present results for two different attack scenarios on a representative distribution grid. We used a modified version of the IEEE-123 node test primary feeder network\footnote{\url{https://cmte.ieee.org/pes-testfeeders/resources/}} shown in \cref{fig:network_diagram}, with SMOs present at 85 out of the 123 total nodes. We added distributed generation (PV, batteries, diesel generators) throughout the network, with the largest amount of generation concentrated at five primary feeder (SMO) nodes numbered 25, 40, 67, 81, and 94. We synthetically generated flexibility bids for all SMOs by assuming that they could provide up to 30\% flexibility around their baseline power injections \cite{Olsen2014Demand2020}. We used 5-minute real-time market LMPs from CAISO and assumed the Q-LMP to be 10\% of the P-LMP \cite{federal_energy_regulatory_commission_payment_2014}. The originally unbalanced feeder was converted to an equivalent balanced 3-phase model by (i) assuming all switches to be at their normal positions, (ii) converting single-phase spot loads to 3-phase, (iii) assuming cables to be 3-phase transposed, (iv) converting configurations 1 thru 12 to symmetric matrices and (v) modeling shunt capacitors as 3-phase reactive power generators \cite{Haider2020TowardGrids}. A PMO was assumed to be at the slack bus (substation), at 13.2kV, with the SMOs at 4.16kV, and each SMA at 120-240V.

Each SMO has between 3-5 SMAs with the number chosen uniformly at random. Since the injection data in the IEEE-123 model was only available up to the primary feeder node level, we artificially randomly disaggregated the injections at each SMO amongst its SMAs, where each SMA could be either a net load or generator. The flexibility bids for the SM were also randomly generated, allowing each SMA to offer flexibilities of up to $\pm 30\%$ away from their baseline. Thus, the upper and lower limits for the bid flexibilities were set as $\underline{P} = P^0 (1 - \underline{\Delta}), \overline{P} = P^{0} (1 + \overline{\Delta})$, where $\underline{\Delta}, \overline{\Delta} \sim \mathcal{U}[0,0.3]$.
We then performed a co-simulation of both the PM and SM. Detailed results for the SM and PM under the nominal case (without an attack) can be found in a previous study \cite{Nair2022AEdge}. In this paper, we focus specifically on the scenario when an attack occurs, as well as the SM commitment score results. Note that our flexibility bids were purely synthetically created, so the resulting flexible ranges in our simulations may be quite large at times and not realistic in some cases. However, our proposed framework can be generally applied to cases where there is less DER flexibility as well.
\subsection{Commitment reliability score trends}
\begin{figure}
 \centering
     \begin{subfigure}[t]{0.49\columnwidth}
         \includegraphics[width=\linewidth]{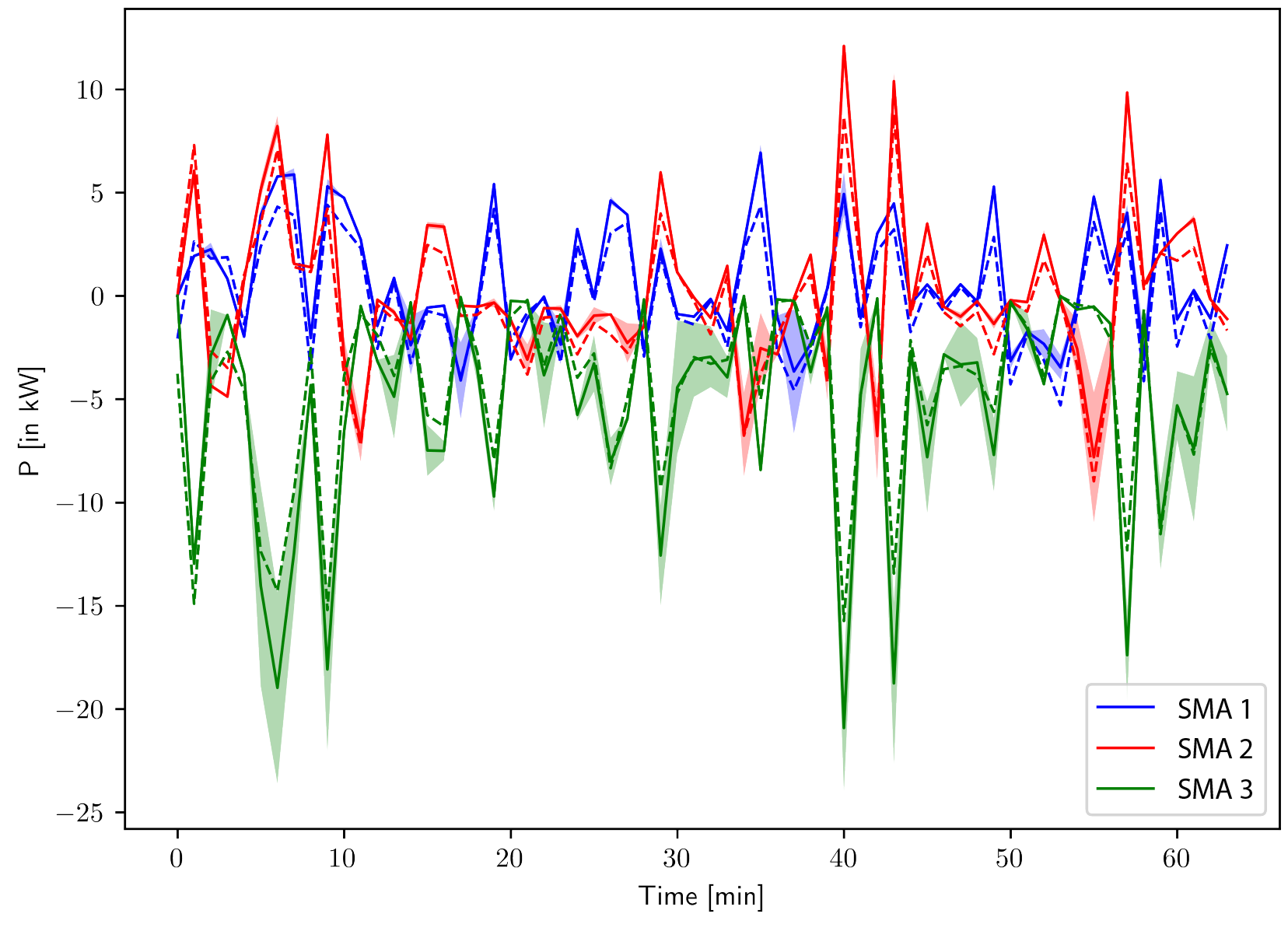}
         \caption{Schedules and responses. \label{fig:sma_injections}}
     \end{subfigure}
     \begin{subfigure}[t]{0.49\columnwidth}
         \includegraphics[width=\linewidth]{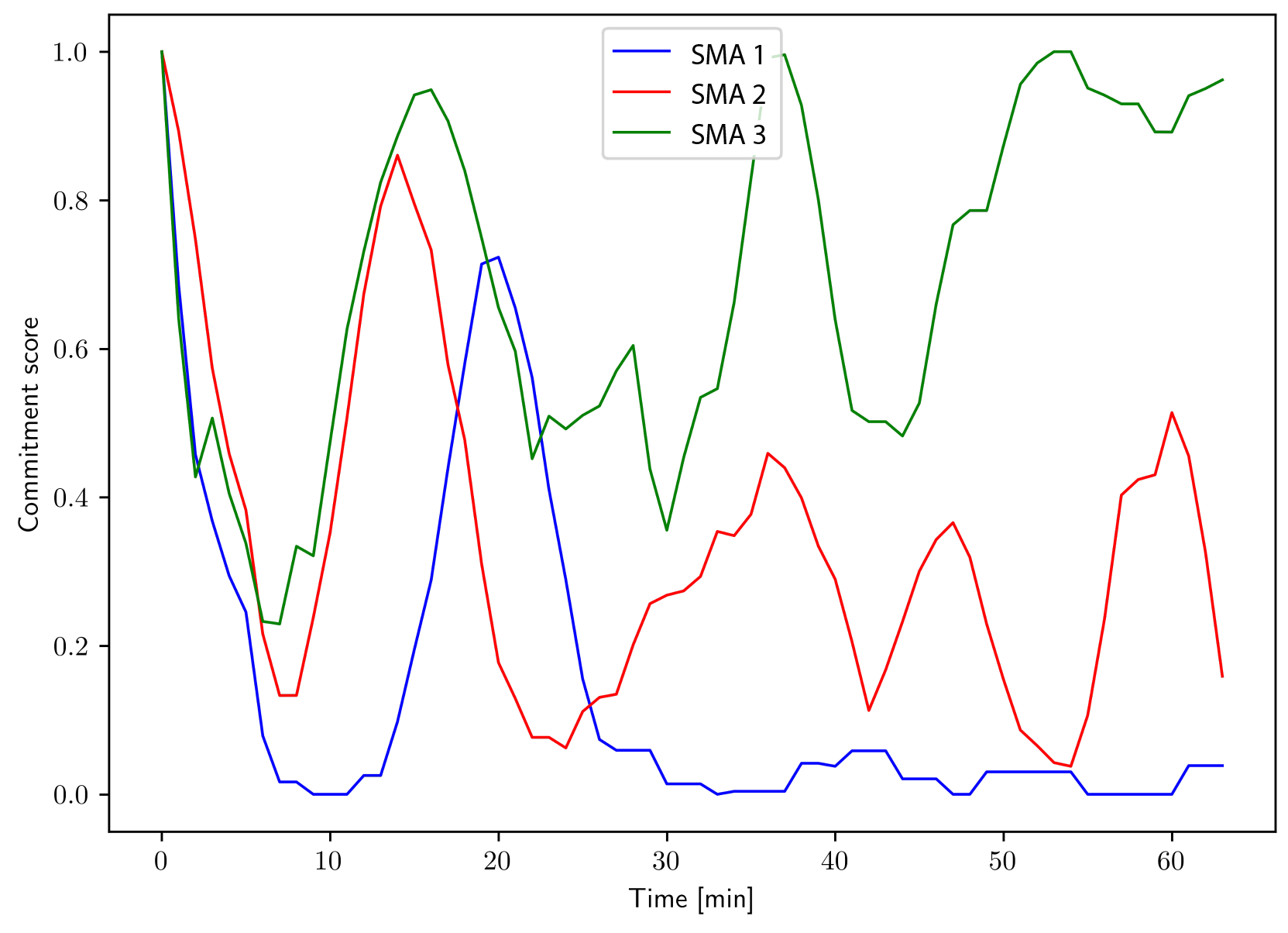}
         \caption{Commitment scores.\label{fig:commit_sma}}
     \end{subfigure}
     \caption{Trends in SMA active power injections and corresponding commitment scores. The solid lines are the nominal schedules resulting from the market clearing, while the shaded region represents the flexible range around the setpoints. The dashed lines are the actual responses of the SMAs.\label{fig:sma_commit_trends}}
\end{figure}
First, we inspect the commitment score results to show that it indeed does reflect the reliability with which an agent can be expected to fulfill their contractual commitments and follow the market schedules. The trends in the commitment scores over 60 minutes for the 3 SMAs connected to SMO 17 are shown in \cref{fig:commit_sma}. By comparing this against the SMAs' corresponding active power injections in \cref{fig:sma_injections}, we see that the commitment score tends to drop whenever the SMAs deviate significantly from their market-assigned schedules and conversely, the score improves when they follow the setpoints more closely. Thus, the PMO and SMOs can use these scores to use more flexibility from DERs with commitment scores closer to 1.

\subsection{Attack simulation and mitigation}
We now simulate two different cyber-physical attack scenarios on this feeder and show how our market structure can successfully mitigate them in a timely manner. Both are disruption attacks where the attacker shuts down one or more of the large DGs in the network. In these simulations, we consider crashed faults that take the generator offline. However, the approach can be easily extended to attack cases where (i) output setpoints of the generators are only partially reduced and/or (ii) loads are increased - both of these would similarly result in a drop in net total generation for the feeder. We only consider a single primary market time step to study the effect of an instantaneous attack. Mitigation can use P dispatch from batteries, P and Q curtailment from flexible loads, limited P dispatch from PV, Q support from smart inverters (connected to PV and batteries), as well as conventional dispatchable fossil fuel sources like diesel generators.

\subsubsection{Small-scale attack \label{sec:small-scale}}
\begin{figure}
\centering
\includegraphics[width=\columnwidth]{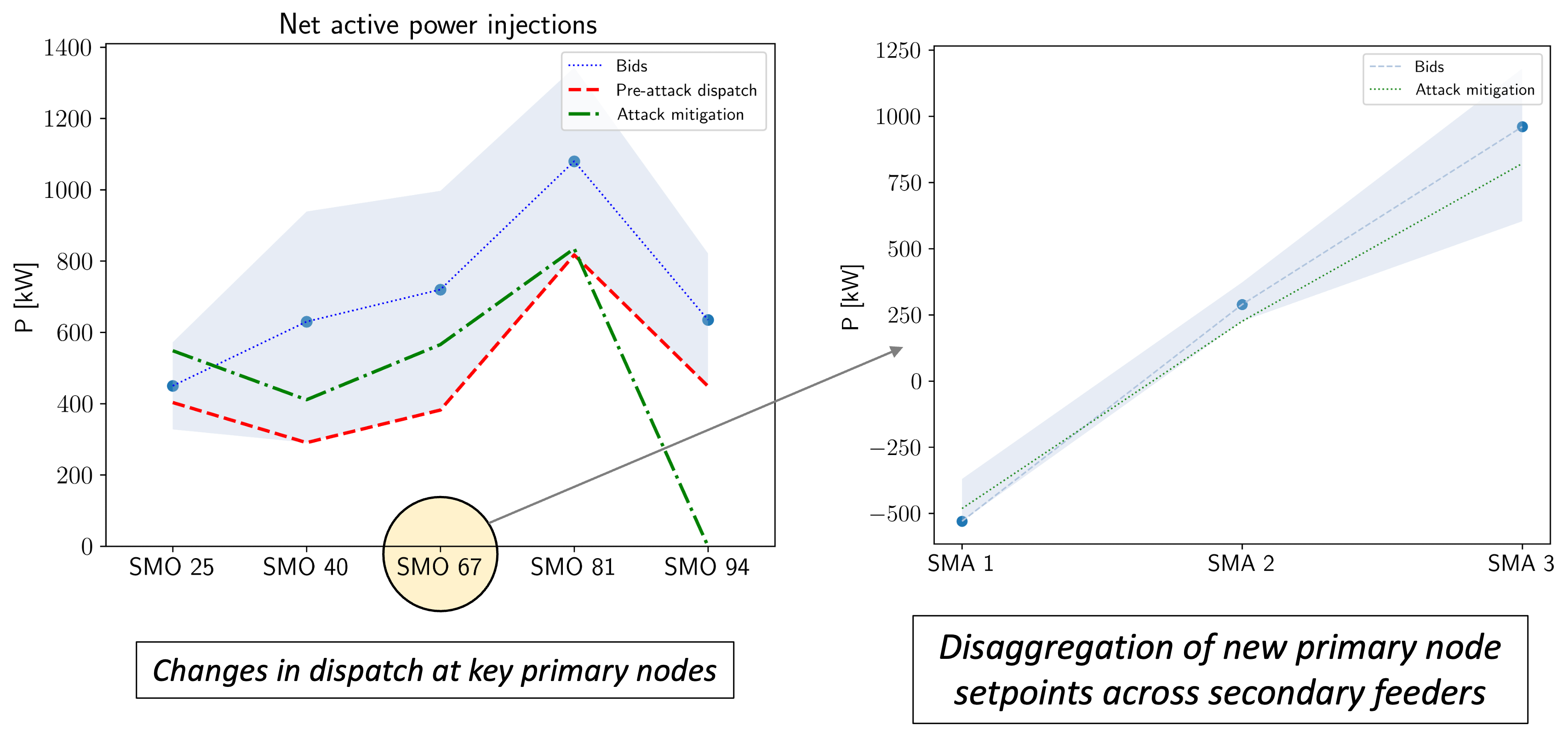}
\caption{Mitigation of small-scale attack. \label{fig:attack_small}}
\end{figure}

Here, only one SMO node 94 is taken offline. Here, we see that the remaining four SMO nodes (25, 40, 67, 81) have more than enough remaining generation capacity to meet the shortfall caused by the attack. Without mitigation, the attack would have resulted in an additional import of about 261 kW from the main grid. However, by utilizing the upward flexibility of remaining SMOs, we're able to fully resolve the attack and bring the total power imported back to pre-attack levels. The left figure in \cref{fig:attack_small} shows the results of the PM dispatch before the attack and after attack mitigation for the five key SMO nodes of interest. The plot also shows the SMO's bids into the PM, with the dashed blue line being the baseline injection bid and the blue-shaded region representing the upward/downward flexibility around it. The right figure shows the results of the SM re-dispatch (after the attack mitigation and PM re-dispatch) for SMO 67 as an example. It disaggregates the new setpoints among its three SMAs, with SMA 1 being a net load while SMAs 2 and 3 are net generators.

\subsubsection{Large-scale attack \label{sec:large-scale}}
\begin{figure}
\centering
\includegraphics[width=\columnwidth]{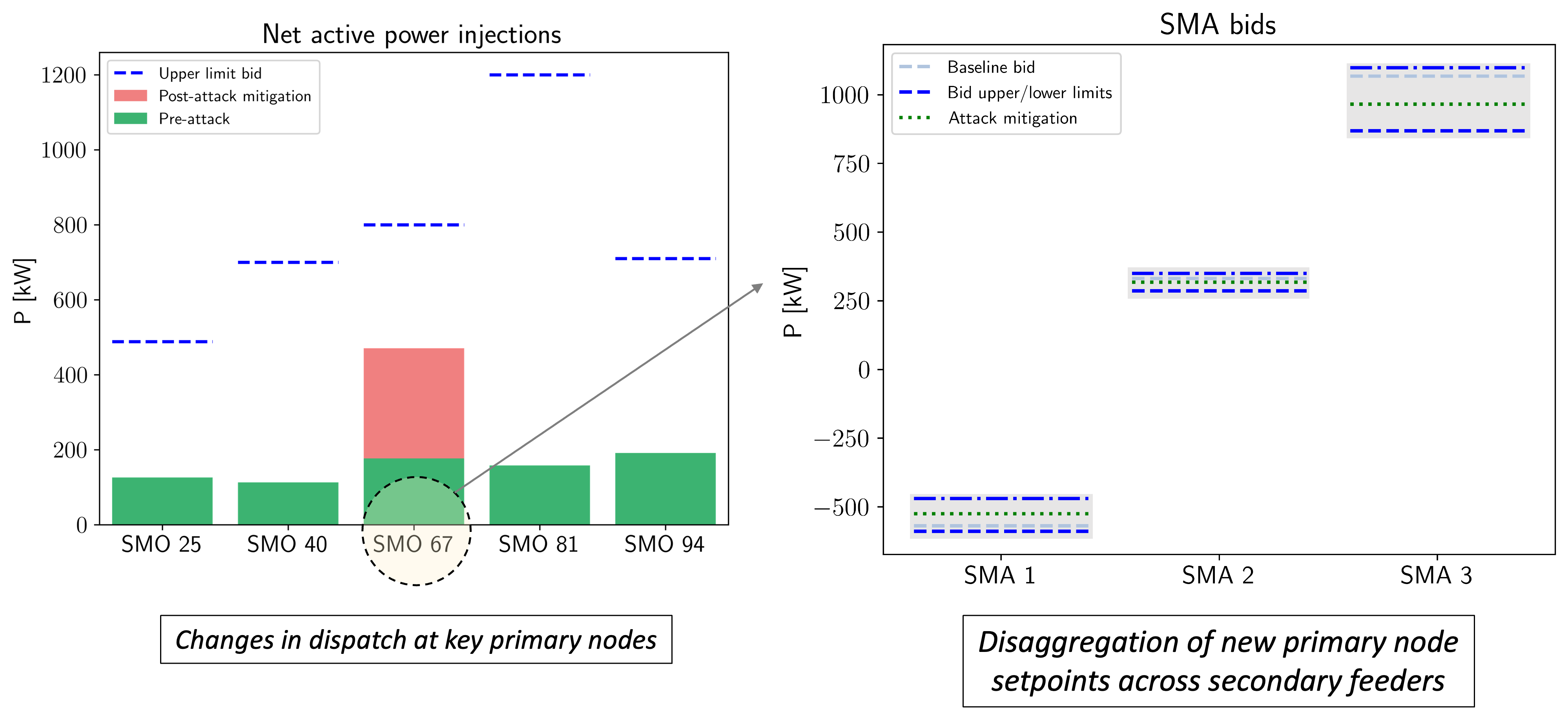}
\caption{Mitigation of large-scale attack. \label{fig:attack_large}}
\end{figure}

\begin{figure}
 \centering
     \begin{subfigure}[t]{0.49\columnwidth}
         \includegraphics[width=\columnwidth]{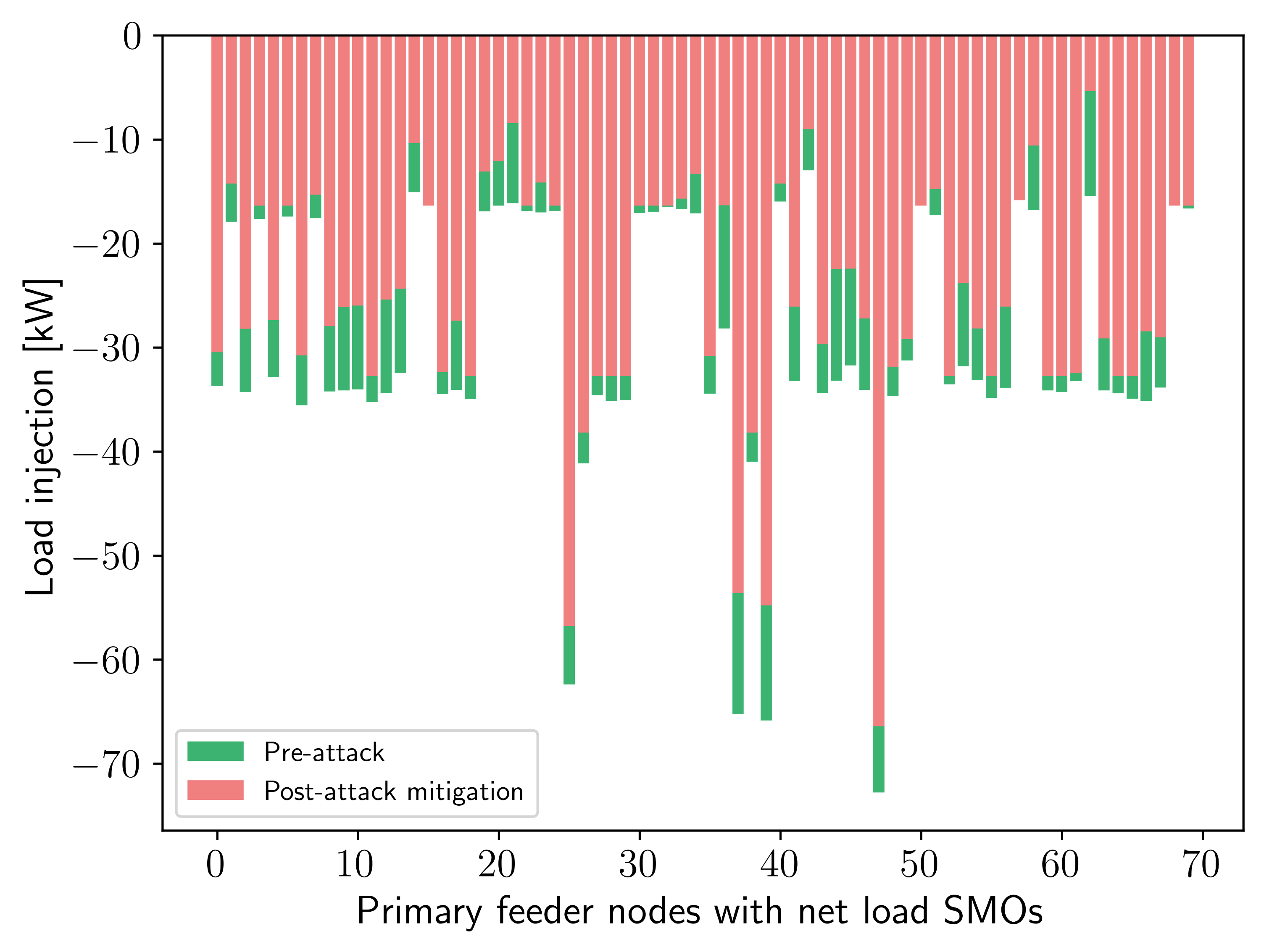}
      \caption{Flexible load curtailment. \label{fig:attack_large_loads}}
     \end{subfigure}
     \begin{subfigure}[t]{0.49\columnwidth}
         \includegraphics[width=\columnwidth]{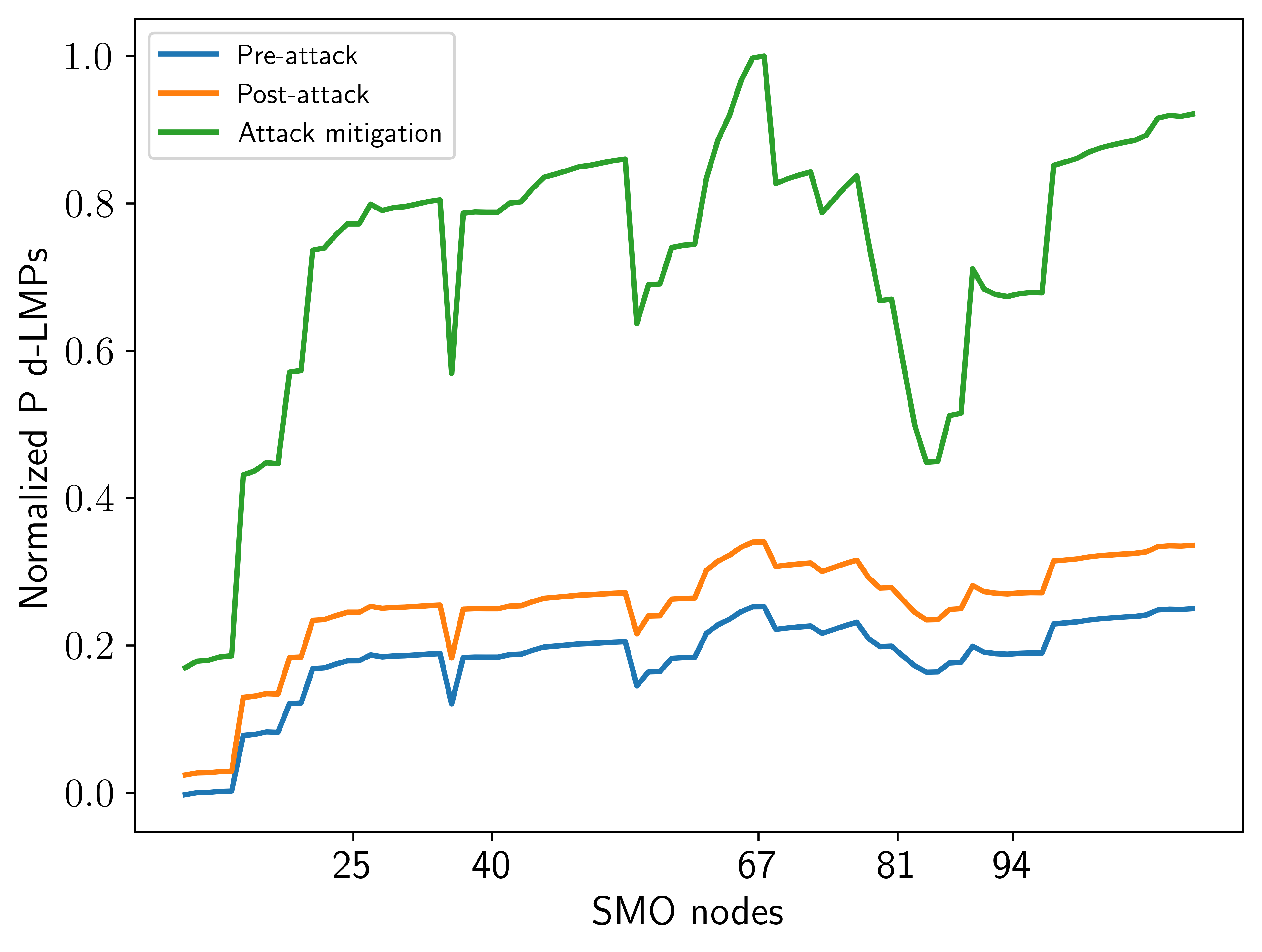}
    \caption{Normalized P d-LMPs. \label{fig:attack_large_prices}}
     \end{subfigure}
     \caption{Effects of large-scale attack and mitigation on flexible load curtailment and nodal prices at SMO nodes. \label{fig:attack_large_loads_prices}}
\end{figure}

In this case, we attack four SMOs at nodes 25, 40, 81, and 94. The only remaining SMO with significant generation capability is at node 67. In \cref{fig:attack_large}, we see how the SM and PM are re-dispatched to mitigate the attack. Similar to the small attack, we leverage the upward generation flexibility of the remaining SMO 67 to increase its output injection after attack mitigation, while the net injections for all the other four attacked SMOs drop to zero as seen in the left plot. The right plot shows the new SMA schedules resulting from the revised SM clearing.
However, one important difference here is that due to the larger scale of the attack, re-dispatching the generator SMOs is no longer sufficient to fully meet the shortfall. Furthermore, as seen in \cref{fig:attack_large}, we are not able to utilize all the upward flexibility of the remaining online SMO 67 since its dispatch is limited by power flow constraints, on nodal voltages and line currents in particular. Thus, we also need to perform some shifting and curtailment of high-wattage flexible loads. These could include electric vehicles, and thermostatically controlled loads like heating, ventilation, and cooling (HVAC), and water heaters. In addition, it could also involve some discharging of battery storage systems to reduce net load. The distribution of net load reductions across the remaining SMOs is shown in \cref{fig:attack_large_loads}, with a total decrease of around $14 \%$ as seen in \cref{tab:metrics}. From \cref{tab:metrics} we also see that the attack would have potentially increased the power import from the transmission grid by over $37 \%$, but the combination of increased local generation and load curtailment helps keep the imported amount almost the same as before.

We also compared the normalized d-LMPs for active power before and after the attack, as well as post-attack mitigation, shown in \cref{fig:attack_large_prices}. As intuitively expected, we see that nodal prices increase throughout the grid after the attack and rise even further after implementing the attack mitigation steps, indicating that the loss of some local generation makes it more expensive to satisfy network constraints and results in sub-optimal solutions. The pre-attack and post-attack prices have nearly the same spatial profile across all the SMO nodes, with the post-attack values essentially being higher by an offset. This makes sense because the d-LMP variations between nodes are influenced by congestion on lines. In the attack case without mitigation, the shortfall caused by the attack would've been compensated for entirely by importing extra power from the grid, and thus the relative congestion variation over the rest of the network remains largely unchanged. The price trends after attack mitigation look more different since the changes in power flow and congestion (resulting from the PM re-dispatch) are not uniform throughout the network. Notably, we see that the prices are significantly more volatile, especially around the nodes affected by the attack. The prices also peak at node 67 - this makes sense since it has the highest increase in injections after attack mitigation, which in turn worsens congestion in lines connected to it.

\subsection{Metrics, cost, economic and distributional impacts}
In our simulations, we find that attack mitigation comes at the expense of increased operational costs for the PMO since it needs to dispatch more expensive local resources to a greater extent, rather than importing cheaper power from the main grid (at the LMP rate). The PMOs and SMOs also need to adequately compensate agents for the critical flexibility they provide, including payments to flexible customers whose loads are curtailed. As shown in \cref{tab:metrics}, the attack increases the system operating costs by around $7\%$, and the mitigation steps raise the cost by over $31\%$, both relative to the pre-attack case. However, the PMO could recoup these costs through other revenue streams and cost savings. For example, the transmission system operators may compensate PMOs for locally containing attacks. Being able to leverage local DER flexibility through markets could also reduce the amount of auxiliary backup generation that the PMO needs to maintain, and/or lower the reserves it may have to otherwise procure from capacity or ancillary service markets. The PMO in turn could also redistribute some of these benefits amongst the SMOs and SMAs.

Another important consideration is the impact of our mitigation approach on the different market participants, i.e., the SMOs and SMAs themselves. The objective function update rules from \cref{eq:update1,eq:update2} generally imply that these local resources will be compensated less per unit (kW or kVAR) of grid support they provide, either in terms of load flexibility or generation dispatch. It may also lead to significant load shifting and curtailment in order to meet grid objectives, which can reduce the overall utility of end-users. We also need to more carefully study the distributional impacts of such methods since they may end up disproportionately negatively impacting certain groups of customers or prosumers, which could in turn have important implications for energy affordability, equity, and fairness.

One possible approach to account for such considerations could be to introduce an additional social welfare (SW) prioritization parameter $\kappa>0$ that allows the PMO to control the tradeoff between meeting global grid objectives (e.g. mitigating attacks quickly, minimizing extra transmission imports, maintaining grid frequency, etc.) versus more local objectives of maximizing total welfare of market agents (e.g. serving more load, compensating DERs more fairly, etc.). We can implement this by modifying the coefficient updates to:
$$
\Delta_{\alpha} = \Delta_{\beta} = \kappa \cdot \frac{|P_{PCC}|}{|P_{PCC}^{\prime}|}, \; \Delta_{\xi} = \left(\frac{1}{\kappa}\right) \cdot \frac{|P_{PCC}^{\prime}|}{|P_{PCC}|}\label{eq:update3}
$$
Thus, choosing $\kappa > 1$ would imply that the PMO focuses relatively more on maximizing social welfare and minimizing local impacts on agents and DERs, potentially at the cost of global grid-related goals by importing some more power.
\begin{table}
\centering
% \begin{tabular}{@{}llll@{}}
\caption{Summary of key system metrics for large-scale attack scenario. \label{tab:metrics}}
\begin{tabular}{@{}p{2.5cm}lll@{}}
\toprule
                                                    & \textbf{Pre-attack} & \textbf{Post-attack} & \textbf{Attack mitigation} \\ \midrule
\textbf{Power import from main grid {[}kW{]}} & 1,325               & 1,821 (+37.4\%)      & 1,328                           \\
\textbf{Total cost {[}\${]}}                        & 10,752              & 11,500 (+7\%)        & 14,156 (+31.7\%)                \\
\textbf{Total load {[}kW{]}}                        & 2,064               & 2,023 (-0.02\%)      & 1,775 (-14\%)                   \\ \bottomrule
\end{tabular}
\end{table}

\section{Conclusions and future work \label{sec:conc}}
In this paper, we applied a hierarchical retail electricity market structure to help detect and mitigate cyber-physical attacks on a distribution grid. A new commitment score metric helps quantify the reliability of agents in following market schedules and can be used by market operators during dispatch. We utilize markets at two different levels, with the SM coordinating multiple secondary feeders using decentralized optimization, and the PM coordinating nodes across the primary feeder using distributed optimization. We develop a new technique to re-dispatch DERs efficiently by modifying various cost coefficients and parameters in the objective function. This approach does not require the market operator to have full visibility over the agents and minimizes the communication needed, allowing for faster responses to the attack. Through numerical simulations on a modified IEEE 123-node test feeder, we validate that our method is able to mitigate both small and large-scale attacks, with little to no extra power import from the transmission system. We also analyzed the effects of attack mitigation on other system metrics like operating costs and considered its broader impacts on social welfare.

% We do not explicitly study deception or disclosure attacks in this paper. However, we'll comment on some potential ways to extend our approach to these cases in \cref{sec:results}.
\subsection{Observations, limitations and future work}

So far, we have only considered attacks that result in an actual drop of net generation in the network and manifest as a change in feeder power import that can be detected at the substation or PCC. These represent disruption (or denial of service) attacks that affect the actual availability of power. Our current simplistic and intuitive approach is able to resolve these effectively, can be easily integrated into a market, and implemented in a purely distributed fashion. However, we recognize that there are other possible types of attack scenarios where our approach may fail, and may require more sophisticated techniques. In the future, we plan to improve this mechanism to successfully detect and mitigate more complex attacks, including disclosure and deception attacks. For instance, to augment our current threshold-based detection approach that uses only the PCC injection values, we can leverage some other recent attack detection approaches as proposed in \cite{sia2018pmu,sia2023phasor}, that also utilize additional data from phasor measurement units (PMU) throughout the network. In addition, we saw in \cref{sec:large-scale} that mitigating the large-scale attack also required extensive load curtailment. However, our approach may be insufficient in case of even larger attacks and if there aren't enough reserves in terms of load or generation flexibility. Here, other mitigation strategies may be needed in addition to modifying cost coefficients, if we don't want to resort to ancillary services or importing more power from the transmission grid.

For future work, we plan to address these limitations and also consider other possible model extensions and enhancements. We will study more distributed attacks where larger numbers of smaller DERs are simultaneously attacked. In doing so, we plan to replicate some attack models proposed in the literature, that show how attacking large numbers of grid-edge IoT devices can potentially destabilize the grid \cite{soltan2018blackiot_new,shekari2022madiot_new}. In addition to the commitment reliability score, we can also consider other metrics that measure the trustability of IoT devices \cite{Sarker2023ResiliencyIoTs}. In addition to cyber-attacks, our framework could also potentially be used to reduce the impacts of other disturbances like extreme weather events.

Another thrust of our future work will be to make our studies more accurate and realistic. We will extend our analysis to more general meshed and unbalanced networks using a current injection-based power flow model \cite{Ferro2020ANetworks}. To create more realistic DR scenarios, we hope to include more physical and operational constraints on how DERs can provide flexibility. While inverter-based resources can be controlled on fast timescales, ramp rate limits may be needed for diesel generators. While charging and discharging batteries, in addition to the capacity and power limits, other factors may also be relevant such as the battery health, cycling degradation, etc. There may also be extra constraints on how different kinds of flexible loads can be operated, with prior works distinguishing between curtailable, shiftable, and non-interruptible loads \cite{Petersen2013AGrids}. Finally, we plan to study the effects of large-scale distribution grid attacks and our mitigation strategies, on the broader transmission system in order to understand their implications for grid frequency and stability.

% using the well-known Kundur 2-area system model, where our distribution grid in area 2 is attacked. We show the effects of the attack and the mitigation strategy on grid frequency in figure ..... Thus, attack mitigation is essential to prevent distribution grid attacks from affecting the larger grid and maintain grid stability.

% Using blockchain and distributed ledgers to improve security, transparency and trustability, and execute smart contracts.

\bibliographystyle{IEEEtran}
\bibliography{refs_mendeley,refs_manual}

\section{Appendix \label{sec:appendix}}

\subsection{Hierarchical multiobjective optimization \label{sec:hierarchy_app}}
For the SM clearing, we use a hierarchical approach to solve the multiobjective optimization problem in \cref{eq:opt}. Here, the SMO optimizes one objective at a time in descending rank and places additional constraints on the degradation of prior objectives while optimizing subsequent ones, as in \cref{eq:degrade}. Here, $\epsilon$ is a hyperparameter that controls the extent to which previous objective(s) are allowed to be degraded while considering the next objective(s). This is chosen by the decision maker (i.e. the SMO) according to their preference - we used a value of $\epsilon = 0.01$ (i.e. a 1\% degradation) in our simulations.
\begin{align}
    \min_{\vec{S}_j} & \; F_i = \sum_{j \in \mathcal{N}} f_{j}^i (\vec{S}_j) \; \forall \; i = 1, 2, 3, 4 \nonumber \\
    \text{s.t.} \; & f_{j}^i(\vec{S}_j) \leq (1 + \epsilon)\sum_{j \in \mathcal{N}}  f_{j}^i (\vec{S}_j^*) = (1 + \epsilon)F_i^*, \label{eq:degrade} \\
    & \forall \; \ell = 1,2,\dots, k-1, \; k>1 \nonumber \\
    & \text{and constraints in eqs.} \; (\ref{eq:Plim}) \; \text{to} \; (\ref{eq:PQbalance}) \nonumber
\end{align}

\subsection{Computation of commitment scores \label{sec:commit_app}}
We describe here the details of computing the commitment reliability score, mentioned in \cref{sec:commit}. From the SM clearing, the SMAs $j$ are directed to keep their net injections within the intervals $[P_j^* - \delta P_j, P_j^* + \delta P_j]$. We first compute the deviations (if any) in their actual responses $\hat{P}_j$ from this range, where $\llbracket \cdot \rrbracket$ denotes the indicator function: 
\begin{align}
    e^{P}_j(t_s) & = \llbracket\hat{P}_j > \overline{P}_j^{*}\rrbracket (\hat{P}_j - \overline{P}_j^{*}) + \llbracket\hat{P}_j < \underline{P}_j^{*}\rrbracket(\underline{P}_j^{*} - \hat{P}_j) \nonumber \\ & + \llbracket\underline{P}_j^{*} \leq \hat{P}_j \leq \overline{P}_j^{*}\rrbracket\max (\hat{P}_j - \overline{P}_j^{*},\underline{P}_j^{*} - \hat{P}_j) \label{eq:commit1}
\end{align}
We then obtain relative deviations by comparing these with the magnitudes of their corresponding baseline setpoints:
\begin{equation}
    \overline{e^{P}_j} (t_s) = \frac{e^{P}_j(t_s)}{|P_j^{*}(t_s)|}, \quad \overline{e^{Q}_j} (t_s) = \frac{e^{Q}_j(t_s)}{|Q_j^{*}(t_s)|} \label{eq:commit2}
\end{equation}
These are then normalized to unit vectors to compare the deviations among all SMAs overseen by the SMO. This allows the SMO to assess their relative performance across SMAs.
\begin{equation}
        \widetilde{\mathbf{e^P}}(t_s) = \frac{\mathbf{\overline{e^P}}(t_s)}{{\|\mathbf{\overline{e^P}}}(t_s)\|}, \; \widetilde{\mathbf{e^Q}}(t_s) = \frac{\mathbf{\overline{e^Q}}(t_s)}{\|\mathbf{\overline{e^Q}}(t_s)\|} \label{eq:commit3}
\end{equation}
The scores are then updated, with the score being increased when SMAs follow their contracts and decreased otherwise:
\begin{align}
C_j(t_s) & = \begin{cases} 1 &\mbox{if } t_s = 0 \\
C_j(t_s - 1) - \frac{\widetilde{e^{P}_j} (t_s) + \widetilde{e^{Q}_j} (t_s)}{2} & \mbox{if } t_s > 0 \end{cases} \label{eq:commit4}
\end{align}
Finally, we perform min-max normalization across all the SMAs' scores to ensure that $0 \leq C_j \leq 1 \; \forall \; \text{SMAs } j$.
$$
    \overline{C}_j = \frac{C_j - \max_j C_j}{\max_j C_j - \min_j C_j}
$$

\subsection{Interface between the SM and PM \label{sec:interface_app}}
Here, we describe the link between the primary and secondary markets. The PM is cleared every 5 minutes while the SM operates at a 1-minute frequency. Thus, before each PM clearing, the SMOs aggregate the schedules of their SMAs' resulting from the most recent prior SM clearing in order to submit their flexibility bid to the PM. All market bidding and clearing for both the SM and PM are based on forecasts (assuming perfect foresight) and for the very next time period.
\begin{align}
    & P^0(t_p) = \sum_{j \in \mathcal{N}}  P_j^{*}(t_p),\; Q^0(t_p) = \sum_{j \in \mathcal{N}}  Q_j^*(t_p) \label{eq:aggregation}  \\
    & \Delta P = \big[\underline{P} = \sum_{j\in \mathcal{N}} P_j - \delta P_j^{*}, \overline{P} =\sum_{j\in \mathcal{N}} P_j^{*} + \delta P_j^{*} \big] \nonumber \\
    &  \Delta Q = \big[\underline{Q} = \sum_{j\in \mathcal{N}} Q_j^{*} - \delta Q_j^{*}, \overline{Q} = \sum_{j\in \mathcal{N}} Q_j^{*} + \delta Q_j^{*}\big] \nonumber
\end{align}

\subsection{Primary market optimization \label{sec:pm_opf_app}}

The detailed optimization problem formulation for the PM is given in \cref{eq:pmo}, where $R$ and $X$ denote the network resistance and reactance matrices respectively, $v$ and $I$ denote the nodal voltage magnitudes and branch currents respectively, and $\mathcal{E}$ denotes the set of all edges in the network.
\begin{align}
    & \underset{y}{\min} \;\; f^{S-W}(y) \label{eq:pmo} \\
    & \text{subject to:} \nonumber\\
	& v_{i} - v_{k} = \left(R_{ki}^{2} + X_{ki}^{2}\right) |I_{ki}|^2 - 2 \left(R_{ki} P_{ki} + X_{ki} Q_{ki}\right)\nonumber\\
	& P_{i}^{G} - P_{i}^{L} = -P_{ki} + R_{ki} |I_{ki}|^2 + \sum_{k:\left(ik\right) \in \mathcal{E}} P_{ik}\nonumber\\
	& Q_{i}^{G} - Q_{i}^{L} = -Q_{ki} + X_{ki} |I_{ki}|^2 + \sum_{k:\left(ik\right) \in \mathcal{E}} Q_{ik}\nonumber\\ 
  	& P_{ki}^{2} + Q_{ki}^{2} \leq \overline{S}_{ki}^{2}, \; P_{ki}^{2} + Q_{ki}^{2} \leq v_{i} |I_{ki}|^2, \;\underline{v}_i \leq v_i \leq \overline{v}_i  \nonumber\\
	& \underline{P}_i^{G,L} \leq P_i^{G,L} \leq \overline{P}_i^{G,L}, \; \underline{Q}_i^{G,L} \leq Q_i^{G,L} \leq \overline{Q}_i^{G,L} \nonumber
\end{align}

The objective function used is a weighted linear combination of (i) maximizing social welfare in \cref{eq:cost:2}, (ii) minimizing total generation costs in \cref{eq:cost:3} and (iii) minimizing electrical line losses in \cref{eq:cost:4}. The total cost includes paying the locational marginal price (LMP) $\lambda$ for importing power from the transmission grid at the point of common coupling (PCC), as well as the payments to local generator SMOs that provide net positive injections into the PM. The local generation cost coefficients $\alpha_i^P, \alpha_i^Q$ of each SMO $i$ are computed as a weighted average of the SM retail tariffs of its SMAs. Similarly, the disutility coefficients of each SMO (associated with load flexibility) $\beta_i^P, \beta_i^Q$ are given by the weighted mean of the disutility coefficients of their respective SMAs. In the case of the PM, we divide by suitable base values to convert all quantities to per unit (between 0 and 1 p.u.). Thus, it is reasonable to combine all the terms into a single objective function using a simple weighted sum. The parameter $\xi$ controls the tradeoff between penalizing line losses versus optimizing for other objectives.
\begin{align}
    f^{S-W}(y)&=\sum_{i \in \mathcal{N}}\Big[f_{i}^{\text{Load-Disutil}}(y) + f_{i}^{\text{Gen-Cost}}(y)\Big]  \nonumber \\
    & + \xi\Big[\sum_{(ki) \in \mathcal{E}}\!f_{ki}^{\text{Loss}}(y)\Big] \label{eq:cost:1} \\
    f_{i}^{\text{Load-Disutil}}(y)&= \beta_{i}^{P}(P_{i}^{L}-P_{i}^{L0})^2+\beta_{i}^{Q}(Q_{i}-Q_{i}^{L0})^2 \label{eq:cost:2} \\
	f_{i}^{\text{Gen-Cost}}(y)&= \begin{cases}\alpha_{i}^{P} (P_{i}^{G})^2 +\alpha_{i}^{Q} (Q_{i}^{G})^2, & \\
	\lambda_{i}^{P} P_{i}^{G}+\lambda_{i}^{Q} Q_{i}^{G},\text{if } i & \text{ is PCC}
	\end{cases} \label{eq:cost:3}\\
	f_{ki}^{\text{Loss}}(y)&=R_{ki}|I_{ki}|^2 \label{eq:cost:4}
\end{align}

\end{document}